\documentclass[12pt,preprint]{aastex}

\def\lax{{$\mathrel{\hbox{\rlap{\hbox{\lower4pt\hbox{$\sim$}}}\hbox{$<$}}}$}}
\def\gax{{$\mathrel{\hbox{\rlap{\hbox{\lower4pt\hbox{$\sim$}}}\hbox{$>$}}}$}}

\usepackage{color}
\slugcomment{}

\shorttitle{Radiatively inefficient accretion flows induced by gravitational-wave emission 
before massive black hole coalescence}
\shortauthors{Kimitake Hayasaki}

\begin{document}


\title{Radiatively inefficient accretion flows induced by gravitational-wave emission before massive black hole coalescence}

\author{Kimitake Hayasaki}
\affil{Department {of} Astronomy, Kyoto University 
Oiwake-cho, Kitashirakawa, Sakyo-ku, Kyoto 606-8502, {Japan}}
\email{kimi@kusastro.kyoto-u.ac.jp}

\begin{abstract}
We study an accretion flow during the gravitational-wave driven evolution of binary massive black holes. 
After the binary orbit decays due to {an interaction} with a massive circumbinary disk, 
the binary is decoupled from the circumbinary disk 
because the orbital-decay timescale due to emission of gravitational wave 
becomes shorter than the viscous timescale evaluated at the inner edge of circumbinary disk.
During the subsequent evolution, 
the accretion disk, which is truncated at the tidal radius because of the tidal torque, 
also shrinks as the orbital decay.
Assuming that the disk mass changed by this process is all accreted, 
the disk becomes radiatively inefficient 
when the semi-major axis is several hundred Schwarzschild radii.
The high-energy radiations, in spite of a low bolometric luminosity, 
are emitted from an accretion disk around each black hole
long before the black hole coalescence as well as the gravitational wave signals.
The synchrotron process can {notably} produce potentially observable radio emissions 
at large distances if there is a strong, dipole magnetic field around each black hole.
In unequal mass-ratio binaries, step-like light variations are seen in the observed light curve 
because the luminosity is higher and its duration time are shorter 
in the radio emission by the disk around the secondary black hole than those of the primary black hole.
Such a precursor would be unique to not a single black hole system but a binary black hole system, 
and implies that binary black holes finally merge without accretion disks.
\end{abstract}


%%%%%%%%%%%
\section{Introduction}
%%%%%%%%%%%
Astrophysical disks are ubiquitous and sources of active phenomena
in the various system of the Universe; 
the star-compact object systems, star-planet systems, 
active galactic nuclei (AGNs), and so forth.
Although the standard theory of accretion disks\citep{ss73} 
has succeeded in explaining them, there are still many unsolved problems.

High-energy emission from galactic black hole binaries and AGNs
cannot be produced
by the accretion flow{, except for the accretion flow accompanied by a corona,} 
derived from the standard disk theory,
where radiative cooling is so efficient that the thermal energy 
heated up by the viscosity is locally radiated away.
If the radiative cooling becomes inefficient, the thermal energy 
is transported inward with accretion, and then the flow becomes hot. 
Such a flow can produce high-energy emission and is generally 
called as radiatively inefficient accretion flow (RIAF) (Chap.~9 of \citealt{kato08}).

The proto-type model of the RIAF is 
an optically thin advection-dominated accretion flow (ADAF)(\citealt{i77,ny94,ny95,ab95}; see also Chap.~9 of \citealt{kato08} for a review). A low luminosity with high-energy emission 
can be explained by the ADAF with a significantly lower accretion rate than $\dot{M}_{\rm{E}}=L_{\rm{E}}/c^2\sim2\times10^{-2}M_7[M_\odot\rm{yr^{-1}}]$, where $L_{\rm{E}}=4\pi cGM_{\rm{bh}}/\kappa_{\rm{es}}\sim1.3\times10^{45}M_7[\rm{ergs^{-1}}]$, $c$, $G$, $M_7$, and $\kappa_{\rm{es}}$ are the Eddington luminosity, light speed, gravitational constant, black hole mass normalized by $10^7M_\odot$, and electron scattering opacity, respectively. Observed power-low spectra of X-rays from Sgr $A^{*}$ was well reproduced by ADAFs\citep{man97}.

Most galaxies are thought to have massive black holes at their centers \citep{kr95} and
to coevolve with them\citep{mag98,fm00,geb00}. 
Galaxy merger leads to the mass inflow to the central region 
and then a nucleus of the merged galaxy is activated 
and black hole grows by gas accretion. During a sequence of processes, binary massive  black holes with a sub-parsec scale 
separation are inevitably formed before two black holes merge by emitting 
gravitational radiation\citep{bege80,es05,do07,may07}.
By interaction with surrounding stars 
{(\citealt{mm05} and references therein; \citealt{be09})}, 
%(\citealt{mm05} and references therein; \citealt{se07}; \citealt{matsu07}; \citealt{be09}), 
gaseous disks\citep{iv99, go00, armi1, haya09, cu09, ha09}, 
other massive black holes\citep{iwa06}, and infalling dwarf galaxies\citep{matsui09},
the binary evolves towards coalescence within Hubble time, 
{although whether the binary finally coalesces still remains uncertainty theoretically (e.g., \citealt{lo09})
and observationally.}
Then, gravitational radiation emitted {prior to coalescence} is detectable with the {\it Laser Interferometer Space Antenna} ({\it LISA}) and Pulser Timing Arrays(PTAs).

{
An electromagnetic counterpart of gravitational wave signal plays a key role 
%in localizing the source and 
in determining the redshift of the source.
By combining the redshift with its luminosity distance the gravitational wave determines,
one can obtain the distance-redshift relation, which is one of best observational probes of the dark energy\citep{schutz86,hh05}. }
Some types of the electromagnetic counterparts have been studied
such as afterglows\citep{mp05,ro10,cor10,tk10} and precursors as the prompt emissions\citep{cp09,bode10,nl10} 
and periodic emissions\citep{haya07,haya08,bo08,mm08,khato09}.

In this Letter, we investigate RIAFs triggered by 
the rapid orbital decay due to emission of gravitational wave. 
The letter is organized as follows. 
The evolution of binary massive black holes interacting 
with a massive circumbinary disk are briefly described in Section~2.
In Section~3, we derive the possible accretion flows and 
their luminosities during the gravitational-wave driven evolution.
Section~4 is devoted to summary and discussion.

%%%%%%%%%%%%%%%%%%%%%%%%%%%%%%%%%
%%%%%%%               Figure 1
%%%%%%%%%%%%%%%%%%%%%%%%%%%%%%%%%
\begin{figure}[!ht]
{
\resizebox{\hsize}{!}{
\includegraphics[width=86mm]{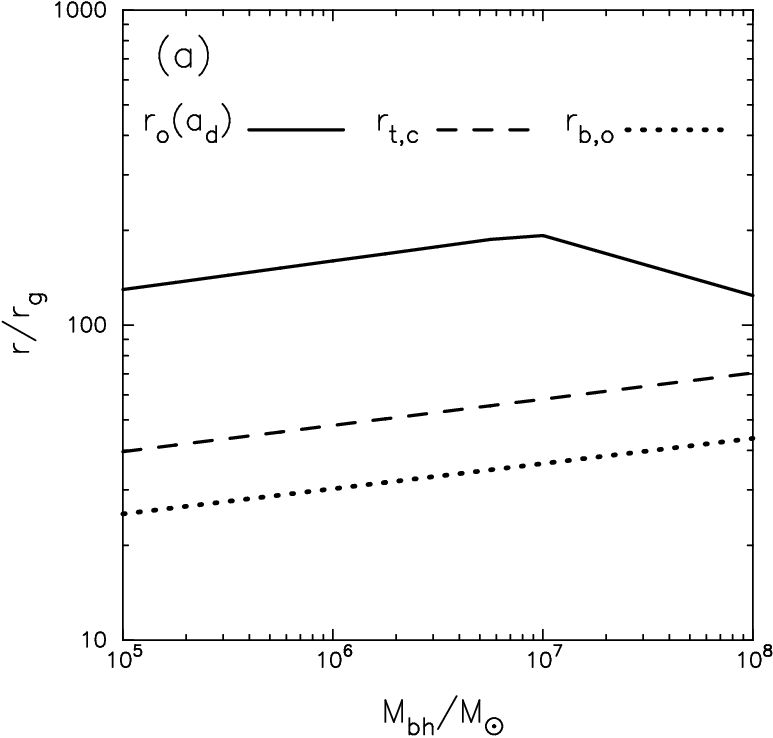}
\includegraphics[width=86mm]{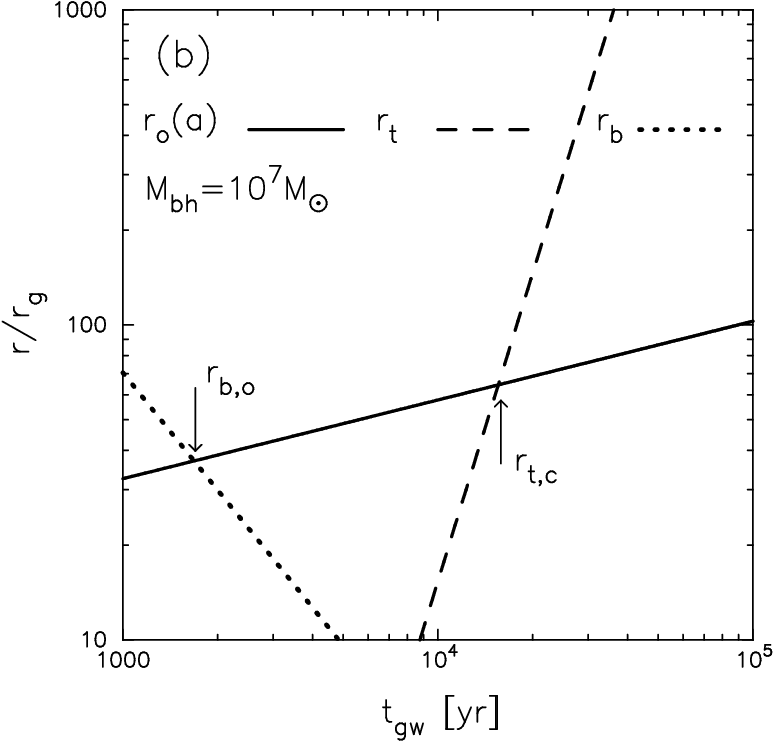}}
}
\caption{
{(a)}Black-hole mass dependence of characteristic radii of the accretion disk 
around the primary black hole {in the binary with $q=0.1$}.
%in equal-mass binary black holes. 
{
The solid line represents the tidal radius 
when the binary is decoupled from the circumbinary disk.
The dashed line shows the critical transition radius, $r_{\rm{t,c}}$, 
where the disk-outer edge corresponds to
the transition radius, $r_{\rm{t}}$, from the standard disk to the RIAF.
The dotted line shows the radius, $r_{\rm{b,o}}$, where the disk-outer edge corresponds to
the boundary, $r_{\rm{b}}$, between the inner, radiation-pressure dominated region 
and the middle, gas-pressure dominated region. (b)Evolution of characteristic radii for $M_{\rm{bh}}=10^7M_\odot$.
The solid line, dashed line, and dotted line 
represent the disk-outer edge, $r_{\rm{o}}(a)$, $r_{\rm{t}}$, and $r_{\rm{b}}$, respectively.  
The intersection between the solid line and the dashed line shows  $r_{\rm{t,c}}$, whereas
the one between the solid line and the dotted line
shows $r_{\rm{b,o}}$.
}
}
\label{fig:adini}
\end{figure}
%%%%%%%%%%%%%%%%%%%%%%%%%%%%%%%%%
%%%%%%%               Figure 2
%%%%%%%%%%%%%%%%%%%%%%%%%%%%%%%%%
\begin{figure}[!ht]
\resizebox{\hsize}{!}{
\includegraphics[width=20mm]{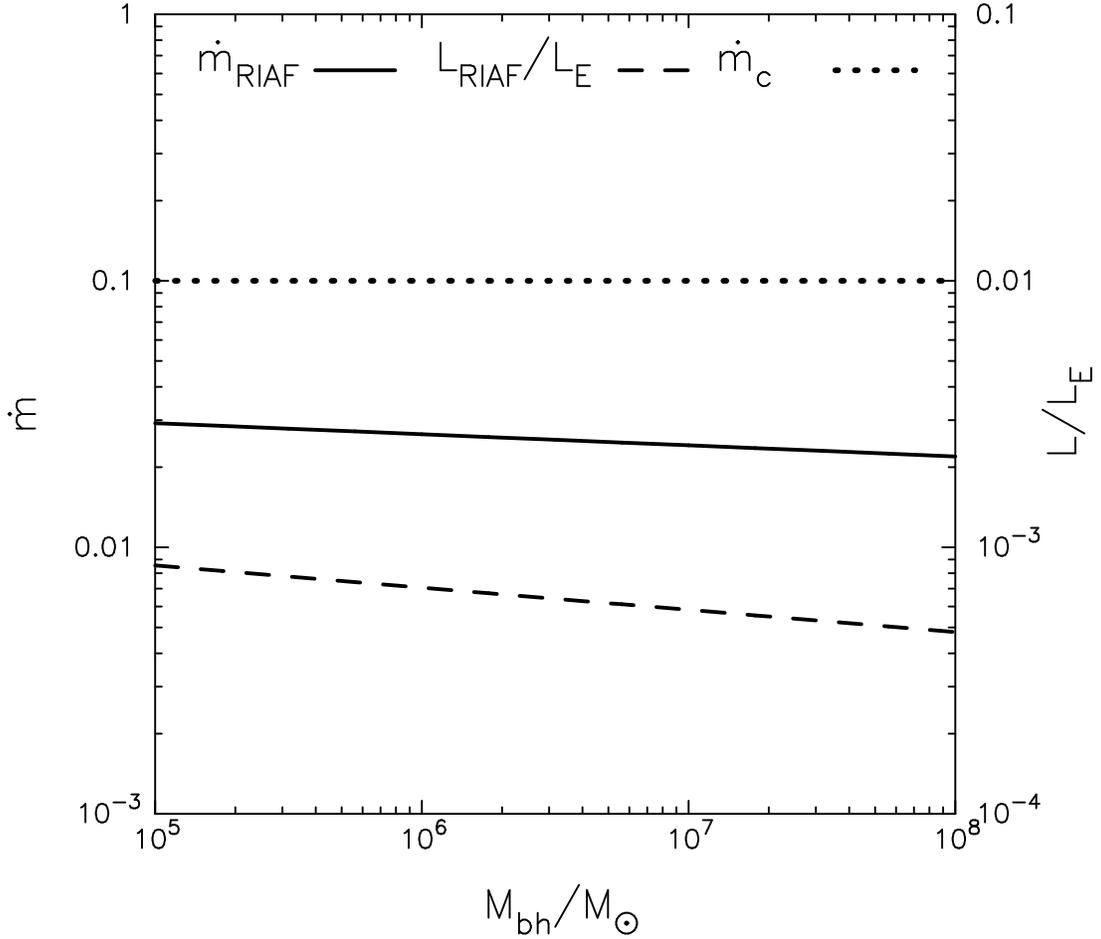}
%\\\includegraphics[width=76mm]{adafevomd.ps}
}
\caption{Black-hole mass dependence of the normalized accretion rate 
and corresponding normalized luminosity {in the binary with $q=0.1$}.
% {in equal-mass binary black holes.
{
The normalized accretion rate, $\dot{m}_{\rm{RIAF}}$, 
is evaluated at the critical transition radius, $r_{\rm{t,c}}$.}
The solid line and dashed line show $\dot{m}_{\rm{RIAF}}$ and 
corresponding normalized luminosity, $L_{\rm{RIAF}}/L_{\rm{E}}$, respectively.
The dotted line shows the normalized critical accretion rate, $\dot{m}_{\rm{c}}$.
%The dash-dotted line shows both the normalized accretion rate of standard, radiation-pressure dominated disk, $\dot{m}_{\rm{SD}}$, at the transition radius and the corresponding normalized luminosity.
}
\label{fig:mdotlumi}
\end{figure}
%%%%%%%%%%%%%%%%%%%%%%%%%%%%%%%%%

%%%%%%%%%%%%%%%%%%%%%%%%%%%%%%%
\section{Final Parsec Evolution of Binary Massive Black Holes}
%%%%%%%%%%%%%%%%%%%%%%%%%%%%%%%
We first briefly describe the evolution of binary massive black holes interacting 
with a massive circumbinary disk 
in the framework of coevolution of massive black holes 
and their host galaxies. The detailed description can be seen in \cite{haya10}.

Binary black holes are considered mainly to evolve via three stages \citep{bege80}.
Firstly, each of black holes sinks independently towards the center of the common 
gravitational potential due to the dynamical friction with neighboring stars.  
When the separation between two black holes
becomes less than $1\,\rm{pc}$ or so, 
angular momentum loss by the dynamical friction 
slows down due to the loss-cone effect and a massive hard binary is formed.
This is the second stage. The binary hardens at the radius
where the kinetic energy per unit mass of the star 
with the velocity dispersion equals to the binding energy per
unit mass of binary black holes\citep{q96}.
Its hardening radius is defined as
$a_{\rm{h}}\sim8.5\times10^{-1}[\rm{pc}](q/(1+q)^2)M_7^{1-2/\beta_2}$,
where $q=M_{\rm{s}}/M_{\rm{p}}$ is the ratio of primary black-hole mass, $M_{\rm{p}}$, 
to secondary black-hole mass, $M_{\rm{s}}$.
Here, it is assumed that 
the total black-hole mass is tightly correlated with the velocity dispersion:
$M_7=\beta_1\left(\sigma_*/200\rm{kms^{-1}}\right)^{\beta_{2}}$
with $\beta_1=16.6$ and $\beta_2=4.86$, where $\sigma_*$ shows the one-dimensional velocity dispersion of stars\citep{mm05}.
Finally, the semi-major axis of the binary decreases to the radius at which the gravitational 
radiation dominates, and then a pair of black holes merge into a single, more massive one.

%%%%%%%%%%%%%%%%%
\subsection{Triple-Disk Evolution}
%%%%%%%%%%%%%%%%%
The circumbinary disk would be formed around hardening of the binary. 
The inner edge of circumbinary disk, $r_{\rm{in}}$, 
is then defined as
$r_{\rm{in}}\approx2a\sim1.8[{\rm{pc}}](q/(1+q)^2)M_7^{1-2/\beta_2}(a/a_{\rm{h}})$
, where $a$ is the semi-major axis of binary.
For simplicity, the circumbinary disk is assumed to be 
a steady, axisymmetric, geometrically thin, self-regulated, self-gravitating 
but non-fragmenting with a Keplerian rotation and accretion rate, 
$\dot{M}_{\rm{acc}}=(\eta/\epsilon)\dot{M}_{\rm{Edd}}$, 
with the Eddington ratio, $\eta$, mass-to-energy conversion efficiency, $\epsilon$, and 
{Shakura-Sunyaev type viscosity parameter, $\alpha_{\rm{sg}}$}.

The circumbinary disk and binary exchanges the energy and angular momentum 
through the tidal/resonant interaction, which leads to the orbital decay of the binary.
On the other hand, the gas at the inner edge of circumbinary disk overflows onto the central binary\citep{haya07}. Then, an accretion disk is formed around each black hole\citep{haya08}. The mass transfer therefore adds its angular momentum to the binary via accretion disks\citep{haya09}.  In a steady state, the mass transfer rate equals to $\dot{M}_{\rm{acc}}$. 
Since it is much smaller than the critical transfer rate defined by equation (41) of \citet{haya09}, {the additional torque to the binary by the mass transfer} can be neglected.
The orbital-decay timescale is then given as
%\begin{eqnarray}
$t_{\rm{gas}}\sim 3.1\times10^{8}[{\rm{yr}}](q/(1+q)^2)\eta_{0.1}^{-1}\epsilon_{0.1}$,
%\label{tc0}
%\end{eqnarray}
where $\eta_{0.1}=\eta/0.1$ and $\epsilon_{0.1}=\epsilon/0.1$. 
{The total mass supplied to} the central binary during the orbital decay is obtained as
\begin{eqnarray}
M_{\rm{sup}}=\dot{M}_{\rm{acc}}
%\times 
t_{\rm{gas}}\sim6.2\times10^6[M_\odot]
M_7\frac{q}{(1+q)^2}.
\label{mdup}
\end{eqnarray}
{$M_{\rm{sup}}$ thus gives an upper limit of mass of two accretion disks.}
%{During the binary evolution, the total mass of two accretion disks must thus be less than $M_{\rm{sup}}$.}

%%%%%%%%%%%%%%%%%%%%%%%%%
\subsubsection{Decoupling of Circumbinary Disk}
%%%%%%%%%%%%%%%%%%%%%%%%%
The orbital-decay timescale of the binary by gravitational-wave emission 
can be written by \cite{p64} as
\begin{eqnarray}
&&
t_\mathrm{gw}=\biggr|\frac{a}{\dot{a}}\biggr|_{\rm{gw}}
=\frac{5}{{8}}
\left(\frac{a}{r_{\rm{g}}}\right)^4
\frac{r_{\rm{g}}}{c}
\frac{(1+q)^2}{q}
\frac{(1-e^2)^{7/2}}{f(e)},
\label{tgr}
\end{eqnarray}
where $e$ is the orbital eccentricity, $f(e)=(1+73e^2/24+37e^4/96)$ is a function of the orbital eccentricity, 
and {$r_{\rm{g}}=2GM_{\rm{bh}}/c^2$ is the Schwarzschild radius of total black hole mass. 
For simplicity, the orbital eccentricity set to be zero, unless otherwise noted, in what follows.}

{
As the binary evolves, the circumbinary disk behaves as the standard disk in the inner region 
for $M_{\rm{bh}}\lesssim10^7M_\odot$, whereas the disk still remains to be self-gravitating 
for $M_{\rm{bh}}\gtrsim10^7M_\odot$.
When the orbital-decay timescale is shorter than the viscous timescale 
measured at the inner edge of the circumbinary disk, 
the binary is decoupled from the circumbinary disk.
For the self-gravitating disk, the decoupling radius can be defined as $a_{\rm{d}}/r_{\rm{g}}=\left[(8/5)(ct_{\rm{vis,g}}/r_{\rm{g}})(q/(1+q)^2)\right]^{2/7}$
, where $t_{\rm{vis,g}}\sim5.6\times10^{4}[\rm{yr}](\alpha_{\rm{sg}}/0.06)^{-1/3}\eta_{0.1}^{2/3}\epsilon_{0.1}^{2/3}M_7^{1/3}$
%for the self-gravitating disk and
is the viscous timescale measured at $a=r_{\rm{g}}$\citep{haya10}.
For the standard disk, $a_{\rm{d}}/r_{\rm{g}}=[(8/5)(ct_{\rm{vis,g}}/r_{\rm{g}})(q/(1+q)^2)]^{4/11}$, 
where $t_{\rm{vis,g}}\sim5.9\times10^{2}[\rm{yr}](\alpha_{\rm{SS}}/0.1)^{-1}\eta_{0.1}^{-1/4}\epsilon_{0.1}^{1/4}M_7^{5/4}$.
After the binary is decoupled from the circumbinary disk,  
the system consists of only an accretion disk around each black hole.
}
%, and the orbital eccentricity gets rapidly close to zero with time.

%%%%%%%%%%%%%%%%%%%%%%
\section{Subsequent Double-Disk Evolution}
%%%%%%%%%%%%%%%%%%%%%%
For simplicity, we focus on the accretion disk around the primary black hole in what follows.
Assuming that the binary has a circular orbit, 
the disk outer edge can be defined by the tidal radius
as
{
$r_{{\rm{o}}}(a)\simeq0.9R_1\sim4.2\times10^{-1}\left(M_{\rm{p}}/M_{\rm{bh}}\right)^{1/3}a$,
}
where $M_{\rm{bh}}=M_{\rm{p}}+M_{\rm{s}}$ and $R_1$ shows the inner Roche radius
of the primary black hole
%, which is approximately written as {$R_1\approx4.6\times10^{-1}a/(q+1)^{1/3}$}
for $q\gtrsim0.1$(e.g., \citealt{fr02}).
The logarithmic differentiation of the tidal radius is given by
\begin{eqnarray}
\frac{\dot{r}_{{\rm{o}}}(a)}{r_{{\rm{o}}}(a)}=\frac{\dot{a}}{a}+
\frac{1}{3}\frac{q}{1+q}\left[\frac{\dot{M}_{\rm{p}}}{M_{\rm{p}}}-\frac{\dot{M}_{\rm{s}}}{M_{\rm{s}}}\right]
\approx\frac{\dot{a}}{a},
\label{dotroche}
\end{eqnarray}
where $\dot{M}_{\rm{p}}$ and $\dot{M}_{\rm{s}}$ are the growth rate of primary black hole and 
that of secondary black hole, respectively. 
$\dot{M}_{\rm{p}}/M_{\rm{p}}\ll1$ and $\dot{M}_{\rm{s}}/M_{\rm{s}}\ll1$ are held during the evolution.

The accretion disk, which is truncated at the tidal radius because of its tidal torque, shrinks with
the orbital decay due to gravitational-wave emission.
The disk mass lying between $r+\Delta{r}$ and $\Delta{r}$
can be then estimated as
\begin{eqnarray}
\Delta{M}=2\pi\int_{r}^{r+\Delta{r}}\Sigma(r)rdr
\approx
2\pi{r}\Sigma(r)\Delta{r}.
\label{adm}
\end{eqnarray}
We assume that the disk mass changed by shrinking of the tidal radius 
steadily accretes onto the black hole without being emitted as jet or wind.
Combining equation~(\ref{adm}) with equations~(\ref{tgr}) and (\ref{dotroche}), 
we obtain the accretion rate normalized by $\dot{M}_{\rm{E}}$:
\begin{eqnarray}
&&
\dot{m}=
\frac{2\pi{r^2}\Sigma}{\dot{M}_{\rm{E}}}\frac{\dot{r}}{r}
=\frac{2\pi{r^2}\Sigma(r)}{t_{\rm{gw}}}\frac{1}{\dot{M}_{\rm{E}}}.
\label{mdota}
\end{eqnarray}

Following \cite{ss73}, the standard disk structure 
has {three distinct regions for a given accretion rate 
due to the source of opacity and pressure. 
In the inner region, the pressure and opacity of the disk 
are dominated by the radiation pressure 
and electron scattering, respectively.
The pressure and opacity are dominated by the
gas pressure and opacity of the electron scattering, respectively, 
in the middle region, whereas they are dominated by 
the gas pressure and opacity of the free-free absorption, respectively, 
in the outer region.}
At the present case, the outer region is neglected because the disk-outer edge 
is always less than the boundary between the middle region and the outer region.
The surface density in the inner region and middle region 
is written by \cite{kato08} as
\begin{eqnarray}
\Sigma(r)=
\left\{\begin{array}{cl}
\Sigma_{\rm{in}}(r/r_{\rm{g}})^{3/2}\dot{m}^{-1}
&  r \le r_{\rm{b}}\\
\Sigma_{\rm{mid}}(r/r_{\rm{g}})^{-3/5}\dot{m}^{3/5}
&  r \ge r_{\rm{b}}
\end{array},
\right.
\label{sigma}
\end{eqnarray}
where {$\Sigma_{\rm{in}}=1.0\times10^{3}\alpha_{0.1}^{-1}(1+q)^{3/2}f^{-1}[\rm{g/cm^2}]$}, 
{$\Sigma_{\rm{mid}}=4.3\times10^{31/5}\alpha_{0.1}^{-4/5}M_{7}^{1/5}(1+q)^{-4/5}f^{3/5}[\rm{g/cm^2}]$} with $f=1-\sqrt{3r_{\rm{g}}/r}$, 
and 
\begin{eqnarray}
\frac{r_{\rm{b}}}{r_{\rm{g}}}\approx1.8\times10^{11/7}(\alpha_{0.1}M_7)^{2/21}\dot{m}^{16/21}(1+q)^{-23/21}
\end{eqnarray}
is the boundary between the inner region and the middle region.
%{ The viscous timescale of accretion disk is given by
%\begin{eqnarray}
%t_{\rm{vis}}=\frac{\sqrt{2}}{2}\frac{1}{\alpha_{\rm{SS}}}\frac{1}{(1+q)^{1/2}}\frac{r_{\rm{g}}c}{c_{\rm{s}}^2}\left(\frac{r}{r_{\rm{g}}}\right)^{1/2},
%\end{eqnarray}
%where $c_{\rm{s}}$ is the sound velocity of the disk in the middle region.}

A mass of accretion disk around the primary black hole at the decoupling 
is defined by $M_{\rm{ad}}=\int_{r_{\rm{isco}}}^{r_{\rm{o}}(a_{\rm{d}})}\Sigma(r)rdr$, 
where $r_{\rm{isco}}=3r_{\rm{g}}/(1+q)$ 
shows the radius of innermost stable circular orbit around the Schwarzschild black hole.
Since $M_{\rm{ad}}$ is
%with equations~(\ref{ro}) and (\ref{sigma}) as 
an increasing function of black hole mass and 
$M_{\rm{ad}}/M_{\rm{bh}}\lesssim10^{-3}$ for $M_{\rm{bh}}=10^{8}M_\odot$ and $q=1$, 
$M_{\rm{ad}}$ is always less than $M_{\rm{sup}}$ over the whole mass range.

From equations~(\ref{mdota}) and (\ref{sigma}), we finally obtain the normalized accretion rate as
\begin{eqnarray}
\dot{m}=
\left\{ \begin{array}{cl}
\left[(2\pi{r}_{\rm{g}}^2\Sigma_{\rm{in}}/t_{\rm{gw}})\dot{M}_{\rm{E}}\right]^{1/2}(r_{\rm{o}}(a)/r_{\rm{g}})^{7/4}
&  r \le r_{\rm{b,o}}\\
\left[(2\pi{r}_{\rm{g}}^2\Sigma_{\rm{mid}}/t_{\rm{gw}})/\dot{M}_{\rm{E}}\right]^{5/2}(r_{\rm{o}}(a)/r_{\rm{g}})^{7/2}
&  r \ge r_{\rm{b,o}}
\end{array},
\right.
\label{mdot}
\end{eqnarray} 
where {$r_{\rm{b,o}}$} is a newly defined boundary between the inner region and middle region,
which can be written as
{
\begin{eqnarray}
\frac{r_{\rm{b,o}}}{r_{\rm{g}}}=\left[\frac{16\pi}{5}\frac{q}{(1+q)^2}\frac{r_{\rm{g}}c}{\dot{M}_{\rm{E}}}\right]^{8/25}
\frac{\Sigma_{\rm{mid}}^{2/5}}{\Sigma_{\rm{in}}^{2/25}}.
\label{rb}
\end{eqnarray}
}
Note that the entire region of the disk is radiation-pressure dominated when $r_{\rm{o}}(a)\le r_{\rm{b,o}}$.

{
The viscous timescale of the accretion disk around the primary black hole is written by 
$t_{\rm{vis}}\approx\sqrt{GM_{\rm{bh}}r/(1+q)}/(\alpha_{\rm{SS}}c_{\rm{s}}^2$), 
where $c_{\rm{s}}$ is the sound velocity of the middle region of the disk.
The ratio of $t_{\rm{vis}}$ to $t_{\rm{gw}}$ is then given by
\begin{eqnarray}
\frac{t_{\rm{vis}}}{t_{\rm{gw}}}\sim(1+q)\left(\frac{r}{r_{\rm{o}}(a)}\right)^{7/5},
\label{tvistgw}
\end{eqnarray}
where equation (\ref{mdot}) is substituted in the derivation process.
Note that this ratio is independent of the viscosity parameter and black hole mass.  
The assumption that the disk is the steady state during the gravitational-wave driven evolution
is thus justified in most of region of the disk because of $t_{\rm{vis}}\lesssim t_{\rm{gw}}$.
From equation (\ref{tvistgw}), $t_{\rm{vis}}$ is slightly longer than $t_{\rm{gw}}$ at the disk-outer edge.
If the outer region of the disk is, however, hotter by the tidal dissipation, 
$t_{\rm{vis}}$ would become shorter than $t_{\rm{gw}}$.
In order to confirm whether this solution is realized, 
the time-dependent behavior including the effect of the tidal dissipation 
will be investigated in a subsequent paper.
}

Next, we consider that the disk becomes a bimodal system of standard disk-RIAF, 
where the standard disk in the cool outer region transits to the RIAF in the inner hot region.
We employ a transition radius from the standard disk to the RIAF, 
which is defined by \cite{h96} as 
\begin{eqnarray}
\frac{r_{\rm{t}}}{r_{\rm{g}}}\approx1.1\times10^{-2}\alpha_{0.1}^4\epsilon_{0.1}^{-2}\dot{m}^{-2}(1+q)^{-1}.
\end{eqnarray} 
The critical transition radius where the disk-outer edge corresponds to the transition radius is then written 
by using equation~(\ref{tgr}) and (\ref{mdot}) as
{
\begin{eqnarray}
\frac{r_{\rm{t,c}}}{r_{\rm{g}}}
\sim9.0\times10^{-1}\alpha_{0.1}^{4}\epsilon_{0.1}^{-2}\frac{q}{(1+q)^{43/18}}
\left(\frac{r_{\rm{g}}c\Sigma_{\rm{mid}}}{\dot{M}_{\rm{E}}}\right)^{5/12}.
\label{rt}
\end{eqnarray}
}
%where equation~(\ref{tgr}) and (\ref{mdot}) were used.
Note that the disk becomes the bimodal system when $r_{\rm{o}}(a)\le r_{\rm{t,c}}$.
%Note that the entire region of the disk is the RIAF at $r_{\rm{o}}(a)=r_{\rm{t,c}}$.

Fig.~\ref{fig:adini} shows the mass dependence of characteristic radii 
of the accretion disk in the binary with $q=0.1$, 
and their evolution for $M_{\rm{bh}}=10^7M_\odot$.
All the radii are normalized by $r_{\rm{g}}$.
In panel (a),
the solid line is the tidal radius when the binary is decoupled from the circumbinary disk.
The dashed line and dotted line are $r_{\rm{t,c}}$ and $r_{\rm{b,o}}$, respectively.
Panel (a) shows $r_{\rm{b,o}}<r_{\rm{t,c}}<r_{\rm{o}}(a_{\rm{d}})$ 
in the range from $10^5M_\odot$ to $10^8M_\odot$.
In panel (b),
the solid line, dashed line, and dotted line are 
$r_{\rm{o}}(a)$, $r_{\rm{t}}$, and $r_{\rm{b}}$, respectively.
The intersection between the solid line and the dashed line
represents $r_{\rm{t,c}}$, 
whereas the one between the solid line 
and the dotted line shows $r_{\rm{b,o}}$. 
It is noted from panel (b) that the disk becomes the RIAF
before the disk is radiation-pressure dominated. 

The luminosity of the RIAF normalized by the Eddington value
can be written as(cf. Chap.~9 of \citealt{kato08})
{$L/L_{\rm{E}}\approx\epsilon\left(\dot{m}/\dot{m}_{\rm{c}}\right)\dot{m}$}, 
where $\dot{m}_{\rm{c}}\sim1.0\times10^{-1}\epsilon_{0.1}^{-1}\alpha_{0.1}^{2}$ is the critical accretion rate 
normalized by $\dot{M}_{\rm{E}}$.
{Note that} there are no possible solutions of RIAFs for $\dot{m}>\dot{m}_{\rm{c}}$ \citep{ny95}.
Fig.~\ref{fig:mdotlumi} shows the mass dependence of 
the normalized accretion rate and corresponding normalized luminosity {in the binary with $q=0.1$.}
The solid line and dashed line show the normalized accretion rate {evaluated at the critical transition radius, $\dot{m}_{\rm{RIAF}}$, and corresponding normalized luminosity, $L_{\rm{RIAF}}/L_{\rm{E}}$, respectively. The dotted line shows $\dot{m}_{\rm{c}}$.}
It is clear from the figure that there is a possible solution for the RIAF because of $\dot{m}_c>\dot{m}_{\rm{RIAF}}$.
%While $\dot{m}_{\rm{RIAF}}\sim2-3.0\times10^{-2}$ has little dependence on black hole mass in the range from $10^{5}M_\odot$ to $10^{8}M_\odot$,
{
Both $\dot{m}_{\rm{RIAF}}\sim2.0-3.0\times10^{-2}$ and $L_{\rm{RIAF}}/L_{\rm{E}}\sim5.0-9.0\times10^{-4}$ 
slightly decrease with the black hole mass in the range from $10^{5}M_\odot$ to $10^{8}M_\odot$.
}

The virial temperature evaluated at the critical transition radius, $T_{\rm{vir}}\sim GM_{\rm{bh}}m_{\rm{p}}/r_{\rm{t,c}}$, 
is {$10^{10-11}$}K in the current-mass range.
{Such a high temperature disk can produce the X-ray emissions by the 
inverse Compton scattering and thermal bremsstrahlung, and radio emission 
by the synchrotron emission process. 
The luminosity of the synchrotron emission at the critical transition radius
can be estimated by(Chap.6 of \citealt{rl79})
\begin{eqnarray}
&&
\frac{L_{\rm{sync}}}{L_{\rm{E}}}
=\frac{2}{9}\frac{\sigma_{\rm{T}}}{\dot{M}_{\rm{E}}}\frac{\beta^2}{1-\beta^2}
n_{\rm{e}}r_{\rm{t,c}}^3B^2
\nonumber \\
&&
%\sim4.3\times10^{-11/2}
\sim3.1\times10^{11/2}
\eta_{0.1}
\alpha_{0.1}^{-3/2}
M_7^{-3/2}
(1+q)^{1/4}
%\frac{L_{\rm{RIAF}}}{L_{\rm{E}}}
\left(\frac{r_{\rm{t,c}}}{r_{\rm{g}}}\right)^{-17/4},
%B_1^{2}(1+q)^{-3/4}\left(\frac{r_{\rm{t,c}}}{r_{\rm{g}}}\right)^{7/4}
\end{eqnarray}
where $\beta^2=(r_{\rm{t,c}}/r_{\rm{g}})^{-1}(1+q)^{-1}$, 
the electron number density is evaluated by $n_{\rm{e}}\sim\Sigma(r_{\rm{t,c}})/(m_{\rm{p}}H)$, 
where $H$ is the disk-scale hight, and magnetic field is evaluated by 
$B\simeq2.7\times10^5\sqrt{\eta_{0.1}/M_7}(1+q)^{1/2}(r_{\rm{t,c}}/r_{\rm{g}})^{-3}$, 
based on an assumption of a dipole field with the 
strength of an equipartition value around a non-rotating black hole\citep{df08}.
%of the primary black hole.
%by using equation~(18) of \cite{bode10}
%as $L_{\rm{sync}}/L_{\rm{E}}\approx1.1\times10^{-2}M_7$ in the equal-mass binary.
For the binary with $10^7M_\odot$ and $q=0.1$, $L_{\rm{sync}}/L_{\rm{E}}\sim5.0\times10^{-2}$, which 
is significantly larger than that of other processes in the current case.
This is a promising candidate as an electromagnetic counterpart of gravitational wave emitted at the black hole coalescence.
}
%Such a precursor implies that the accretion disk is drained because of the high energy radiation emitted by the disk prior to coalescence.

{
In an unequal mass-ratio binary, 
the critical transition radius of the disk around the primary black hole 
becomes larger than that of the secondary black hole.
This difference causes step-like light variations in the radio emissions.
Since the duration time approximately equals to the dynamical time evaluated at the critical transition radius,
the difference of duration time can be written by $\Delta{t}_{\rm{duration}}=\sqrt{2}(r_{\rm{g}}/c)(r_{\rm{t,c}}/r_{\rm{g}})^{3/2}(1+q)^{1/2}(1-q^{5/6})$. For the binary with $M_{\rm{bh}}=10^7M_\odot$ and $q=0.1$, the difference of duration time becomes $\Delta_{\rm{duration}}\sim2.0\times10^{-3}$ yr.
On the other hand, the synchrotron-luminosity ratio of the secondary black hole 
to that of the primary black hole is 
%almost inversely proportional to the mass ratio.
%\begin{eqnarray}
%\frac{L_{\rm{sync,s}}}{L_{\rm{sync,p}}}
%=q^{-4/3}\frac{r_{\rm{t,c}}-1/(1+q)}{q^{1/3}r_{\rm{t,c}}-q/(1+q)}
$L_{\rm{sync,s}}/L_{\rm{sync,p}}\simeq q^{-5/3}$. 
%\end{eqnarray}
For the binary with $q=0.1$, the emission from the secondary black hole shows one-order of magnitude higher luminosity 
but shorter duration time than those of the primary black hole. The total light curve would thus exhibit step-like variations.
}
%%%%%%%%%%%%%%%%
\section{Summary\&Discussion}
%%%%%%%%%%%%%%%%
We study radiatively inefficient accretion flows onto each black hole  
during the gravitational-wave emission driven evolution of binary massive black holes 
in the framework of coevolution of massive black holes and their host galaxies.

The binary system initially consists of the accretion disk around each black hole
and the massive circumbinary disk surrounding them.
After the binary is decoupled from the circumbinary disk, 
the outer edge of each accretion disk shrinks with the orbital decay 
due to emission of gravitational wave.
Assuming that the disk mass changed by the rapid orbital decay 
is all accreted, the accretion flow {becomes radiatively inefficient 
when the disk-outer edge is less than the critical transition radius.}

%After two massive black holes merge into a single, more massive one due to the rapid orbital decay by gravitational-wave emission, gas in the decoupled circumbinary disk accretes onto the merged black hole. This produces the afterglow of the gravitational wave emitted at black hole coalescence\citep{mp05,tk10}.For $10^5M_\odot$ black hole, the precursor occurs at $\sim1.1\times10^2$ yr, when the orbital period of binary $\sim1.3\times10^{-3}\rm{yr}$, and the subsequent afterglow occurs at $\sim5.4\times10^2$ yr.The precursor could therefore give an evidence for such a very short orbital period binary.
{
The bolometric luminosity corresponding to 
the accretion rate estimated at the critical radius is $5.0-9.0\times10^{-4}L_{\rm{E}}$ in the range from 
$10^{5}M_\odot$ to $10^8M_\odot$.}
The RIAF produces X-ray emissions by the 
inverse Compton scattering and thermal bremsstrahlung, 
and radio emissions by the synchrotron process.
Assuming that the magnetic field is a dipole field around a non-rotating black hole 
and its strength is the equipartition value, 
The radio emission by the synchrotron process is 
much more luminous than those of other processes 
and exhibits step-like variations by a sum of light curve
from each component of the binary.
Such a precursor implies that each accretion disk are drained 
because of the high energy radiation emitted by each disk prior to coalescence.
%This is a promising, distinctive precursor of massive black hole coalescence at large distances.

{
%Moreover, 
The peak frequency of the synchrotron emission
is given by  $\nu_{c}=3\gamma^2eB\sin\alpha/(4\pi m_{\rm{e}}c)$, and 
$\gamma$, $m_{\rm{e}}$, and $\alpha$ are the Lorenz factor, 
electron mass, and pitch angle between the magnetic field and the electron velocity, respectively.
The peak-frequency ratio of the radiation emitted by the disk around the secondary
black hole to that of the primary black hole
can then be estimated by $\nu_{\rm{c,s}}/\nu_{\rm{c,p}}\simeq q^{-3/2}$ because 
the Lorenz factor and magnetic field strength depend on the mass ratio.
%where $\nu_{c}=3\gamma^2eB\sin\alpha/(4\pi m_{\rm{e}}c)$, and $\gamma$, $m_{\rm{e}}$, and $\alpha$ 
The spectrum is therefore expected to have a double peak when one observe 
in a wide range of radio wavelengths.
}

{
A step-like nature of the light curve and double peak nature of the spectrum 
are unlikely to be seen in emissions from an accretion flow around a single massive black hole, 
and thus gives an unique photon diagnosis of binary black hole coalescences at large distances.
%the coalescence of unequal mass-ratio binary black holes.
} 
%After two massive black holes merge into a single, more massive one due to the rapid orbital decay by gravitational-wave emission, gas in the decoupled circumbinary disk accretes onto the merged black hole. This produces the afterglow of the gravitational wave emitted at black hole coalescence\citep{mp05,tk10}.For $10^5M_\odot$ black hole, the precursor occurs at $\sim1.1\times10^2$ yr, when the orbital period of binary $\sim1.3\times10^{-3}\rm{yr}$, and the subsequent afterglow occurs at $\sim5.4\times10^2$ yr.The precursor could therefore give an evidence for such a very short orbital period binary.

\acknowledgments
K.H. would like to thank anonymous referee for the useful comments and suggestions.
K.H. is also grateful to Loeb Abraham, Shin Mineshige, and Takahiro Tanaka for helpful discussions. 
The calculations reported here were performed using the facility at the Centre for Astrophysics \& Supercomputing at Swinburne University of Technology, Australia and at YITP in Kyoto University. 
This work has been supported by the Grants-in-Aid of the Ministry of Education, Science, Culture, and Sport and Technology (MEXT;  19740100, 18104003, 21540304, 22540243, 22340045).

\end{document}